\documentclass[aps,twocolumn,preprintnumbers,floatfix,groupedaddress]{revtex4}

\usepackage{graphicx}
\usepackage{dcolumn}
\usepackage{bm}
\newcommand{\figwidth}{3.in}
\begin{document}
\title{Heat capacity scaling function for confined superfluids}
\author{Kwangsik Nho} 
\affiliation{Center for Simulational Physics, University of Georgia,
 Athens, Georgia, 30602-2451, U.S.A}
\author{Efstratios   Manousakis}
\affiliation{Department   of  Physics   and  MARTECH,  
Florida  State University,
Tallahassee, Florida 32306, U.S.A and \\
Department of Physics,  University of Athens,
Panepistimiopolis, Zografos, 157 84 Athens, Greece
}
\date{\today}
\begin{abstract}
We study the specific heat scaling function of superfluids confined in
cubic geometry and in parallel-plate (film) geometry with open 
boundary conditions (BC)  along the finite dimensions using Monte Carlo 
simulation. For the case of cubic geometry for the superfluid order parameter 
we apply open BC in all three directions.
We also calculated the specific heat scaling
function for the parallel-plate confinement  using open BC
 along the finite dimension and periodic BC along the other two 
dimensions and we find it to be very close to the earlier calculated using 
Dirichlet instead of open BC. We find that the specific heat scaling function
is significantly different for the two different geometries studied.
In addition, we generally find that the scaling function for a fixed given 
geometry when calculated with open BC is quite close to that calculated 
with Dirichlet BC, while when calculated with periodic BC is quite different. 
Our results for both the scaling functions obtained for the parallel-plate 
geometry and for cubic geometry with open BC along the finite dimensions
are in very good agreement with the recent very high quality experimental
measurements with no free parameters.
\end{abstract}

\maketitle

\section{Introduction}

Thermodynamic quantities such as the specific heat become non-analytic
at a critical point associated with a second order phase transition.
 For a  finite (or confined) system with a finite dimension such as a film
characterized by a length $L$,
close enough to the critical point such that the correlation length 
becomes comparable or larger than
$L$,  such thermodynamic quantities are significantly altered;  the reason 
is that the degrees of freedom of the system are correlated to 
each other over the entire system. 
Examples of such confined 
systems are (a) a film of thickness $L$ where the system is confined 
in one spatial dimension, (b)  a bar-like geometry 
with cross-section $L\times L$ 
and infinite length (such a pore or a wire) where the system 
is confined in its two spatial dimensions
or (c) a cubic geometry of size $L^3$ where the system is finite in all three
dimensions.
For any thermodymamic observable we can define a system specific dimensionless 
quantity, a scaling function\cite{fss}; for example, in the case of the 
specific heat near the critical point and for sufficiently large $L$ we may 
define the following scaling function:
\begin{equation}
 f(x=tL^{1/\nu}) = \frac{c(t,L)-c(0,\infty)}{c(0,L)-c(0,\infty)},
\label{scal1}
\end{equation}
where $t = T/T_{\lambda} -1$ is the reduced temperature 
and $c(t,L)$ is the specific heat for the case of the system
confined within a finite length $L$. 
For a given case of confining geometry and given the
condition which the order parameter satisfies at the boundaries
of the system, as $L$ approaches infinity and $t\to 0$, the scaling function
depends only on the value  of the combination $x=tL^{1\over \nu}$,
namely on the length $L$ in units of the correlation
length $\xi(t) \sim t^{-\nu}$.  A  dimensionless function such as $f(x)$ 
defined by Eq.~\ref{scal1} can be
thought of as a universal scaling function for the specific heat for a
well-defined confining geometry. In other words, the scaling function does
not depend on the microscopic details, but only depends
on the nature of the universality class of the system, the confining
geometry and the boundary conditions which are felt by the order parameter.

In this limit ($t \to 0$ and $L \to \infty$) the scaling function 
is different for the three different cases mentioned 
previously for the following reason:
For a fixed value of $x<<1$ 
and for  any large value of $L$ there is always a sufficiently small 
value of $t$ satisfying the condition where the correlation length 
is much larger than $L$. However, in this limit the case (a) is the case of 
a 2D plane, the case (b) is the case of a 1D line and the case (c) is the
zero-dimensional case. Thus the value $f(x)$ for sufficiently small values of 
$|x|$ should be very different for these three geometries.

Though  earlier experiments  
on  superfluid helium  films of finite  thickness \cite{maps}  
seemed to  confirm the
validity  of the finite-size-scaling(FSS), there were later
experiments\cite{rhee,rheephys} 
where it was shown that the superfluid
density of thick helium films  does not satisfy FSS when
the expected  values of critical  exponents were used.   Similarly, in
measurements  of  the  specific  heat of  helium  in  finite
geometries, other  than the expected values for  the critical exponents
were  found \cite{earlyc}.   

More recent experiments in microgravity environment\cite{chex}
as well as earth bound experiments\cite{mehta1,mehta2} are consistent
with scaling and they
have determined the specific heat scaling function for 
the parallel plate (film) geometry (case (a)) and they are in 
reasonable agreement 
with the  scaling function as was predicted by Monte Carlo 
simulations\cite{sm95,bes} and renormalization group techniques\cite{RG}.
While the specific heat scaling function for case (b) confinement
has been theoretically determined\cite{pore} and it was found to 
be significantly suppressed relative to the case (a) there are so-far no
experimental data to compare. 
More recently, the specific heat scaling 
function for the case (c) has been experimentally 
determined\cite{cubes1,cubes2}.

The main goal of this paper is to present the results of our Monte 
Carlo simulations to determine the
specific heat scaling function for  cubes with open boundary conditions (BC)
in all three directions (confining case (c)). In this case the scaling 
function characterizes the zero-dimensional to three-dimensional transition.
Our results for the scaling function are compared to the very recently
obtained experimental results for specific heat scaling function in 
the case of cubic  confinement\cite{cubes1,cubes2}. We find satisfactory
agreement with no free parameters.
In addition, we present results for the specific heat scaling function
for the parallel plate geometry on lattices of size $L_1\times L_2 \times L$ 
with $L_1=L_2 >> L$ where we have applied periodic BC
along the $L_{1,2}$-directions and open BC along the film-thickness 
direction of size $L$. The latter case was carried out in order to
compare the results for {\it Dirichlet BC} (on the top and bottom of the
film) obtained  earlier\cite{sm95,bes}. In Refs~\cite{sm95,bes} it was 
found that while the results with periodic BC along the film-thickness 
direction  were very different from those obtained with Dirichlet BC, the 
results obtained with Dirichlet BC fit the experimental results with no 
free parameter.   In this paper we find that the scaling function
obtained with open BC along the finite dimension is close to that 
obtained with Dirichlet 
and also fits reasonably well the experimental results obtained by 
the so-called Confined Helium Experiment\cite{chex} (CHEX).

\section{Monte Carlo Calculation}
We have performed a numerical study of the scaling behavior of the
specific heat of $^{4}He$ in a cubic and in a film geometry at temperatures
close to the critical temperature $T_{\lambda}$. 
The superfluid transition of liquid $^{4}He$  belongs to the 
universality class of
the three-dimensional $x-y$ model, thus,
we are going to use this model to compute the specific heat at temperatures
near $T_{\lambda}$ using the cluster Monte-Carlo method \cite{wolff}. 
The $x-y$ model on a lattice is defined as 
\begin{equation}
 H = -J \sum_{\langle i,j \rangle} \vec{s}_{i} \cdot 
     \vec{s}_{j}, 
\label{ham}
\end{equation}
where the summation is over all nearest neighbors,
$\vec s = (\cos\theta, \sin\theta)$ is a two-component vector 
which is constrained to be on the unit circle and $J$ sets the energy scale.

We define the energy density of our model as follows:
\begin{equation}
 E = \langle e \rangle = 
  3 - \frac{1}{V} \left\langle \sum_{\langle i,j \rangle} \vec{s}_{i} \cdot 
     \vec{s}_{j} \right\rangle , \label{ed}
\end{equation}
where $V=L^{3}$ for the cubes and $V=HL^{2}$ for the film geometry.
We have calculated the specific heat using the expression
\begin {equation}
 c = V T^{-2} ( \langle e^{2} \rangle - \langle e \rangle ^{2} ).
\end{equation}
The above thermal averages denoted by the angular brackets are computed
according to 
\begin{eqnarray}
\langle O \rangle = Z^{-1} \int \prod_{i} d\theta_{i} \:  O[\theta]
                     \exp( - \beta {\cal H} ), 
\label{ev}
\end{eqnarray}
where ${\cal H} = H/J$ the energy in units of $J$ and $\beta = J/T$.
$O[\theta]$ denotes the dependence of the physical observable $O$ on the
configuration $\{ \theta_i \}$, and the partition function $Z$ is given by
\begin{eqnarray}
  Z=\int \prod_{i} d\theta_{i} \: \exp( -\beta {\cal H} ).
 \label{z}
\end{eqnarray}

We computed the specific heat $c(T,L)$ of the $x-y$ model as a function of 
temperature on several
cubic lattices $L^3$ (with $L=20, 30, 40, 50$ ). 
Open (free) boundary conditions were applied
in all directions, namely the spins on the surface of the cube
are free to take any value. These spins interact only with the 5 nearest 
neighbors, one in the interior and 4 on the surface of the cube and there is
one missing neighbor.  
We have also calculated the specific 
heat scaling function $f_1(x)$ (to be defined in the following section)
for the case of the parallel plate geometry $L_1 \times L_2 
\times L $ ($L_{1,2}>>L$) using
periodic boundary conditions along the long directions of the film
and open BC along the thickness direction $L$. For this case
we need to take the limit $L_{1,2} \to \infty$ first; in Ref.~\cite{bes} 
it was found that using $L_1=L_2 = 5 L$ was large enough, 
in the sense that systematic errors due to the finite-size 
effects from the fact that $L_{1,2}$ are not infinite are smaller 
than the statistical errors for realistic computational time scales.
The present simulations for films were done on lattices 
$60 \times 60 \times 12$, $70 \times 70 \times 14$, and $80 \times 80 \times \
16$.

We used the Monte Carlo method and in particular
Wolff's cluster algorithm\cite{wolff}.
We carried out of the order of 30,000 MC steps 
for thermalization to obtain equilibrium configurations.
We made of the order of 10,000-50,000 measurements
allowing 500 MC steps between successive measurements to obtain
statistically uncorrelated configurations.

\section{Scaling functions}

\begin{figure}[htp]
\includegraphics[width=\figwidth]{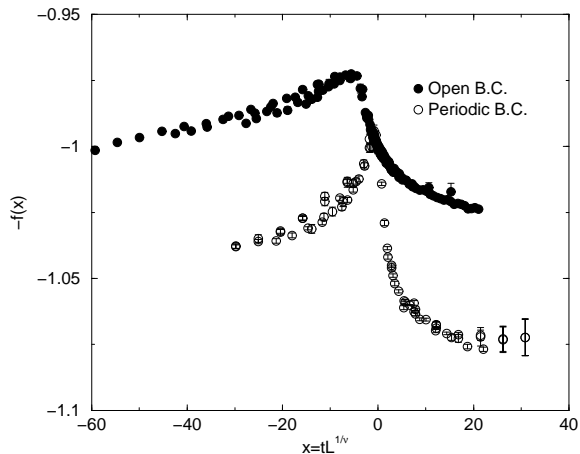}
\caption{The scaling function $-f(x)$ defined by Eq.~\ref{scal1}.}
\label{fi-1}
\end{figure}
The main goal of this paper is to present a calculation of the
specific heat scaling function for the case of cubic confinement and open BC.
In this calculation we have used open BC in all three directions of the
cube.  We found that using open BC the results for the specific
heat scaling were very close to those obtained with Dirichlet BC. 
This will be demonstrated in the next section where we compare the previously
published results\cite{sm95,bes} for films with Dirichlet BC and results
reported in this paper for films with open BC.

One can imagine a number of  different scaling functions for the specific heat.
Any dimensionless combination such as the ratio given by Eq.~\ref{scal1}
can be used as a scaling function. However, the various experimental 
groups have extracted two scaling functions, the so-called $f_1(x)$ and 
$f_2(x)$ with $x=tL^{1\over {\nu}}$. These scaling functions are defined
as follows:
\begin{equation}
c(t,L)-c(t_{0},\infty)= L^{\alpha/\nu}f_{1}(tL^{1/\nu})
\label{scal-f1}
\end{equation}
\begin{equation}
[c(t,\infty)-c(t,L)]t^{\alpha}= f_{2}(tL^{1/\nu})
\label{scal-f2}
\end{equation}
We limit our goal to calculate the specific heat scaling function
for the confined 
geometry and not the bulk critical exponents nor critical amplitude ratios.
We take the values for the bulk critical exponents and the universal 
amplitude ratios as determined experimentally\cite{LPE}. 
Previous MC work such as the work of
Ref.~\cite{sm3d} shows that the critical exponents are within error
bars from the experimental values.
Our approach to use the experimentally determined values of the
critical exponents and amplitude ratios and to determine the scaling
function by applying FSS on the  calculated $c(t,L)$, has no fitting 
parameters and this allows no ambiguity.
Therefore we use $\nu = 0.6709$ as obtained from accurate
experiments\cite{LPE} such as 
the so-called Lambda Point Experiment (LPE), an experiment in microgravity
environment. The hyperscaling relation 
$\alpha = 2-3\nu $ yields
$\alpha/\nu = -0.0189 $, and the correlation length
 $\xi(t)$ = $\xi_{0}^{\pm}|t|^{-\nu}$ becomes equal to the system size $L$ at
the reduced temperature $t_{0}$, i.e., $t_{0}=(\xi_{0}^{+}/L)^{1/\nu}$ with
$\xi_{0}^{+} = 0.498$.

In order to find the universal function $f(x)$ defined by Eq.~\ref{scal1}, 
we need to know
$c(0,\infty)$. We use the bulk values for $c(0,\infty)$ obtained by studying
the finite-size scaling of the specific heat of cubes with periodic
BC\cite{sm3d}. In Fig.~\ref{fi-1} the scaling function $-f(x)$ obtained for 
cubes with open BC in all three directions is compared to that obtained 
with periodic BC\cite{sm3d}.

The scaling function $f_{1}(x)$ (Eq.~\ref{scal-f1}) can be calculated using 
our calculated $c(t,L)$ and 
\begin{equation}
c(t_{0},\infty)=c(0,\infty)+c_{1}^{+}t_{0}^{-\alpha},
\label{eq4}
\end{equation}
where the values of $c(0,\infty)$ and  $c_{1}^{+}$ 
for the $x-y$ model are obtained from reference\cite{sm3d}.

In order to calculate the universal function $f_{2}(x)$ (Eq.~\ref{scal-f2}), 
we  need to know the bulk specific heat $c(t,\infty)$ also. 
Since we are restricting ourselves to the critical region we
may write the following
\begin{eqnarray}
c(t > 0,\infty) & = & c(0,\infty) + c_{1}^{+}t^{-\alpha}, \label{eq4.a} \\ 
c(t < 0,\infty) & = & c(0,\infty) +  c_{1}^{+}/r |t|^{-\alpha}, \label{eq4.b}
 \\ 
r & = & {{c_1^{+}} \over {c_1^{-}}},
\end{eqnarray}
where $r$ is the universal amplitude ratio 
and it is most accurately determined experimentally\cite{LPE,lipachui} 
from the critical properties of bulk helium to be $r = 1.053 (2)$\cite{LPE}.
Inserting Eqs.~\ref{eq4.a} and \ref{eq4.b} into  Eq.~\ref{scal-f2}, we obtain
\begin{eqnarray}
f_{2}(tL^{1/\nu}) = [c(0,\infty)-c(t,L)]t^{\alpha}+c_{1}^{+}, \ \ t>0,\label{eq5a}\\
f_{2}(tL^{1/\nu}) = [c(0,\infty)-c(t,L)]|t|^{\alpha}+c_{1}^{+}/r \ \ t<0 .
\label{eq5b}
\end{eqnarray}
which can be calculated by using our computed $c(t,L)$ and 
the values of $c(0,\infty)$ and  $c_{1}^{+}$ from Ref.\cite{sm3d}.

\section{Films with open boundary conditions}

In Ref.~\cite{sm95,bes} the specific heat scaling function for 
a parallel plate geometry  on lattices of size $L_1\times L_2\times L$ 
with $L_{1}=L_2>>L$ was  calculated.
In Ref.~\cite{sm95,bes} periodic BC
along the $L_{1,2}$-directions and staggered (Dirichlet) BC or periodic 
along the film-thickness direction of size $L$ were applied. 
It was found that while the calculated scaling function for the parallel-plate
geometry using periodic BC along all three directions was 
very different from that obtained with Dirichlet BC along the top and bottom
of the plate and periodic BC along the other two long directions, 
the latter scaling function fits the experimental results with no 
free parameter.  This was explained on the basis that physically
the order parameter along the film thickness vanishes at the boundaries
of the film and therefore Dirichlet BC are more appropriate.

In this paper we have used open BC along the top and the bottom
of the plate, instead of Dirichlet, and periodic BC along the two
long directions of the plate. Since the film terminates on the top and on
the bottom surface, for the pseudospins which belong to these two surfaces 
(in language of the $x-y$ model) there is no neighboring
spins beyond the top and the bottom surface plane of the plate.
Therefore, even if we use open (free) boundary conditions 
this termination acts as ``a zero order parameter 
constraint'' beyond the top and the bottom of the plate.
This implies that these two BC, namely staggered BC and open BC, 
are very similar for thick enough films.

In order to make a direct comparison of our 
calculated $f_{1}(x)$ to the experimental $f_1(x)$, we express
all lattice units in physical units using the following equation\cite{sm95}
\begin{eqnarray}
f_{1}(x)|_{phys} &= \lambda
 f_{1}(x)|_{lattice}, \\
\lambda &\equiv {{V_m k_B} \over {a^3}} (a/\AA)^{-{{\alpha}\over {\nu}}}
\label{lambda}
\end{eqnarray}
where $V_m$ is the molar volume of liquid helium at the lambda point and 
saturated vapor pressure, $k_B$ is Boltzmann's constant and $a$
 the lattice spacing in the $x-y$ model required to make contact with 
the critical behavior of the correlation length in helium.
This prefactor  $\lambda=15.02 J/(K~ mol)$ and it was determined 
in Ref.\cite{sm95}.

\begin{figure}[htp]
\includegraphics[width=\figwidth]{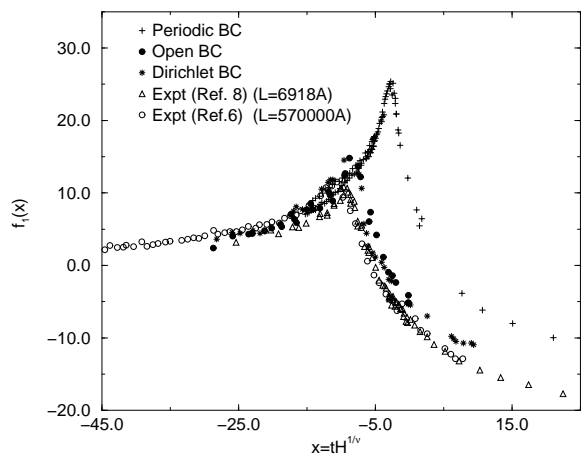}
\caption{Film geometry: 
The computed universal function $f_{1}(x)$ with open BC 
(solid circles) is compared to the previously calculated scaling 
function using Dirichlet BC\cite{sm95}(data shown as stars)
and periodic BC (shown as plus signs) and the experimental 
results of Lipa {\it et al.}\cite{chex} (open circles) and those of 
Mehta {\it et al}.\cite{mehta2} (open triangles).}
\label{fi-7}
\end{figure}

In Fig.~\ref{fi-7} we compare the results for $f_1$ for the case of films
obtained with open BC along the direction of the film thickness to
those obtained earlier\cite{sm95,bes} and to the experimental
results\cite{chex,mehta2}. It is clear that within error bars our results for
the specific heat scaling function are the same for both cases of
BC.

While the scaling function is sensitive to 
boundary conditions, this indicates that it is hard to distinguish 
Dirichlet from
open BC for the specific heat scaling function.
We feel that when we use physical BC the agreement between the 
theoretical results for the specific heat scaling function and 
the experimental results is quite good taking into consideration 
the fact that there is no free parameter.

\section{Cubic confinement}

\begin{figure}[htp]
\includegraphics[width=\figwidth]{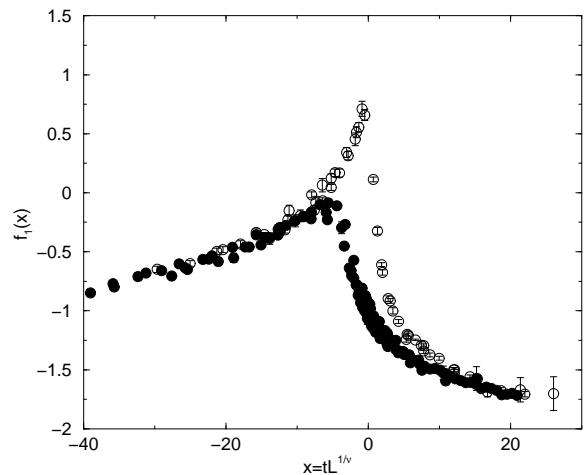}
\caption{The scaling function $f_{1}(x)$ obtained for cubes of size 
$L^3$ with open  (solid circles) and that obtained for cubes 
with periodic BC 
(open circles) are compared.}
\label{fi-2}
\end{figure}

In this section we present the results for the scaling functions
$f_1(x)$ and $f_2(x)$ obtained for cubes of size $L^3$ with $L=20,30,40,50$
using open and periodic BC in all three directions.
As was shown in the previous section  open (free) BC
are similar to using Dirichlet BC and they both express
the physical condition imposed by the confinement or the termination 
of the system.
In Fig.~\ref{fi-2} we compare the scaling function $f_{1}(x)$ 
obtained for open BC with that obtained for periodic BC\cite{sm3d}.
Notice the suppression of $f_1(x)$ when calculated with open BC relative to
the case of periodic BC. This is similar to the case of the parallel-plate
geometry (Fig.~\ref{fi-7}). The scaling functions $f_1(x)$, however,
are very different for cubic and parallel-plate geometry. Notice, for instance,
that for the case of cubic confinement with open BC $f_1(x)$ is negative 
for all values of $x$ something very different of what happens for 
any of the calculated of the experimental 
scaling functions for parallel-plate confinement.
 
\begin{figure}[htp]
\includegraphics[width=\figwidth]{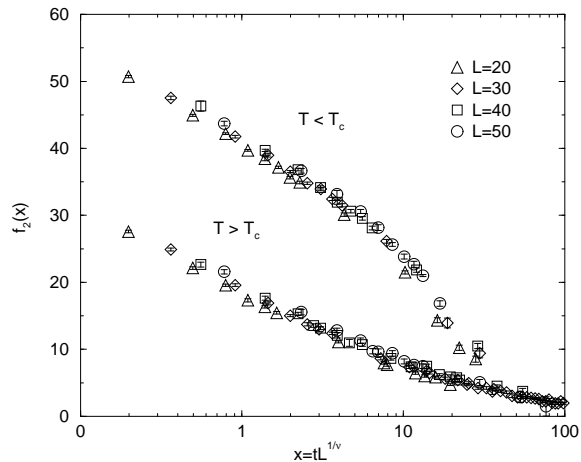}
\caption{The universal function $f_{2}(x)$ for cubes of size $L^3$ 
with open boundary conditions.}
\label{fi-3}
\end{figure}

\begin{figure}[htp]
\includegraphics[width=\figwidth]{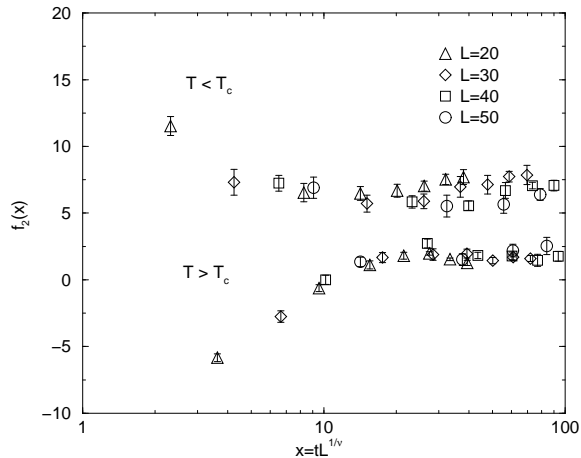}
\caption{The universal function $f_{2}(x)$ for cubes of size $L^3$ with 
periodic boundary conditions}
\label{fi-4}
\end{figure}

\begin{figure}[htp]
\includegraphics[width=\figwidth]{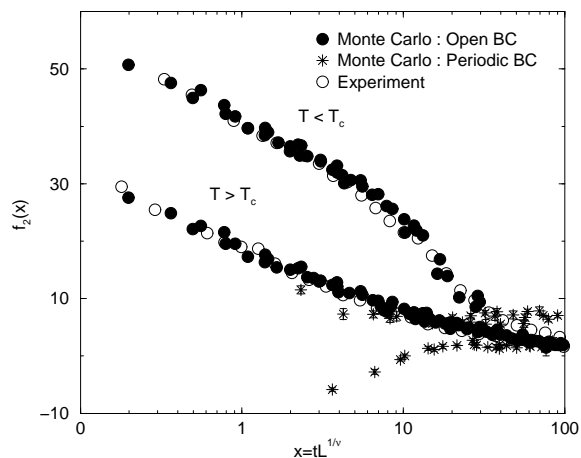}
\caption{The computed universal function $f_{2}(x)$ for open and periodic BC
and for cubes is compared to the experimental results\cite{cubes1,cubes2}.}
\label{fi-6}
\end{figure}

In Fig.~\ref{fi-3} we give the results of  our present Monte 
Carlo calculation of the function $f_2(x)$ for cubes with open BC
using Eqs~\ref{eq5a},\ref{eq5b}. Fig.~\ref{fi-4} shows the results of our
calculation with periodic boundary conditions. 
Fig.~\ref{fi-6} compares the scaling function $f_{2}(x)$ 
obtained for open BC  and for periodic BC. Notice the qualitatively 
different behavior for the same scaling function for the same geometry 
but different  boundary conditions.

Experimentally the universal scaling function $f_2(x)$ for cubic confinement
has just become available\cite{cubes1,cubes2}.
In order to make a direct comparison of our 
calculated scaling function $f_{2}(x)$ to
the experimentally determined, we express
all lattice units in physical units. 
The prefactor is the same as in the case of the function $f_1(x)$:
\begin{eqnarray}
f_{2}(x)|_{phys} &= \lambda
 f_{2}(x)|_{lattice}, 
\end{eqnarray}
where $\lambda$ is the constant given in the previous section 
by Eq.~\ref{lambda} and its numerical value is
 $\lambda=15.02 J/(K~ mol)$.  

In Fig.~\ref{fi-6}  $f_{2}(x)$ obtained from our MC calculation 
is compared with the experimental data\cite{cubes1,cubes2}.
The agreement between the scaling function calculated with open
BC and experiment is quite satisfactory considering the 
fact that there is no free parameters.

\section{Conclusions}

In this paper we have used the $x-y$ model which describes the fluctuations of
the superfluid order parameter near the critical point to calculate 
the scaling functions associated with the specific heat for the case
where the superfluid is confined in a cubic geometry and in parallel-plate
geometry. Both in the theoretical calculations and in the experiments, 
the region very near the superfluid transition is probed such that 
the correlation length
associated with the superfluid order parameter is of the size of the
confining length.

First, we calculated the specific heat scaling function for the case
of parallel plate confining geometry using open boundary conditions
along the top and the bottom surfaces of the film. Our results are very close
to those obtained\cite{sm95,bes} with Dirichlet (staggered BC)
along the top and the bottom surfaces of the film.
Both calculations are in satisfactory agreement with the
experimental results\cite{chex,mehta1,mehta2} while the results 
of earlier calculations using
periodic boundary conditions\cite{sm3d} were found to disagree with the 
experimental scaling function near the superfluid transition.

Just recently, experimental measurements\cite{cubes1,cubes2} 
on superfluid helium confined in cubes became available.
This prompted us to calculate the heat capacity scaling function of
superfluids for cubic confinement. 
When we used open boundary conditions in all three
directions of the cube we find very good agreement between the calculated
and the measured\cite{cubes1,cubes2} scaling functions with no 
adjustable parameter.
On the contrary, if periodic boundary conditions are used at the
boundaries of the cube, which are unphysical boundary conditions for 
a confined system, there is great disagreement between
the calculated and the measured specific heat scaling functions.

\section{Acknowledgments}
This work was supported in part by NASA grants NAG-1773 and NAG-2867 
and by the University
of Athens Office of Scientific Research.
The authors are grateful to M.O. Kimball and F. M. Gasparini who
kindly provided access to their experimental results before
publication and to J. A. Lipa for providing a copy of the most 
recent analysis of the LPE experiment\cite{LPE} before acceptance for
publication.


\begin{thebibliography}{99}

\bibitem{fss} M. E. Fisher and M. N. Barber, Phys. Rev. Lett. {\bf 28} 1516
(1972); M. E. Fisher, Rev. Mod. Phys. {\bf 46} 597 (1974); V. Privman,
Finite Size Scaling and Numerical Simulation of Statistical systems, 
Singapore: World Scientific 1990;
E. Brezin, J. Physique {\bf 43} 15 (1982);
V. Privman, J. Phys. {\bf A23} L711 (1990).
\bibitem{maps} J. Maps and R. B. Hallock, Phys. Rev. Lett {\bf 47} 1533 (1981);
D. J. Bishop and J. D. Reppy, Phys. Rev. Lett. {\bf 40},
1727 (1978).
\bibitem{rhee}I. Rhee, F. M. Gasparini, and D. J. Bishop, Phys. Rev. Lett.
     {\bf 63} 410 (1989).
\bibitem{rheephys} I. Rhee, D. J. Bishop,  and F. M. Gasparini, Physica 
  {\bf B165\&166} 535 (1990).
\bibitem{earlyc} T. Chen and F. M. Gasparini, Phys. Rev. Lett. {\bf 40}
331 (1978); F. M. Gasparini, T. Chen, and B. Bhattacharyya, Phys. Rev. {\bf 23}
5797 (1981).
\bibitem{chex}
J.A. Lipa, D.R. Swanson, J.A. Nissen, Z.K. Geng, P.R. Williamson, 
D.A. Stricker, T.C.P. Chui, U.E. Israelsson and M. Larsen, 
Phys. Rev. Lett. {\bf 84}, 4894 (2000);
and J. Low Temp. Phys. {\bf 113}, 849 (1998).
\bibitem{mehta1}
S. Mehta and F.M. Gasparini, Phys. Rev. Lett. {\bf 78}, 2596 (1997).
\bibitem{mehta2}
S. Mehta, M.O. Kimball and F.M. Gasparini, J. Low Temp. Phys. {\bf 114}, 467 (1999).
\bibitem{sm95} N. Schultka and E. Manousakis, Phys. Rev. Lett. 
{\bf75}, 2710 (1995).
\bibitem{bes} N. Schultka and E. Manousakis, J. Low Temp. Phys. {\bf 109}, 733 (1997).
\bibitem{RG} R. Schmolke, A. Wacker, V. Dohm, and D. Frank, Physica 
 {\bf B165 \& 166} 575 (1990); V. Dohm, Physica Scripta {\bf T49} 46 (1993);
 P. Sutter and V. Dohm, Physica {\bf B194-196} 613 (1994);
W. Huhn and V. Dohm, Phys. Rev. Lett. {\bf 61} 1368 (1988);
M. Krech and S. Dietrich, Phys. Rev. {\bf A46} 1886 (1992).
\bibitem{pore} N. Schultka and E. Manousakis, J. Low Temp. Phys. {\bf 111}, 783 (1998).
\bibitem{cubes1}M. O. Kimball, M. Diaz-Avila and F. M. Gasparini,
Submitted to the editors of the LT23 conference in Hiroshima, Japan.
\bibitem{cubes2}M. O. Kimball and F. M. Gasparini, Private Communication.
\bibitem{wolff}
U. Wolff, Phys. Rev. Lett. {\bf 62}, 361 (1989).
\bibitem{gold}
L.S. Goldner and G. Ahlers, Phys. Rev. B {\bf 45}, 13129 (1992).
\bibitem{sm3d} N. Schultka and E. Manousakis, Phys. Rev. {\bf B52},
7528 (1995).
\bibitem{LPE} 
J. A. Lipa, J. A. Nissen, D. A. Stricker, D. R. Swanson,
T. C. P. Chui, submitted for publication in Phys. Rev. {\bf B};
J. A. Lipa, D. R. Swanson, J. A. Nissen, 
T. C. P. Chui, U. E. Israelsson, Phys. Rev. Lett, {\bf 76}, 944 (1996).
\bibitem{lipachui} J. A. Lipa and T. C. P. Chui, Phys. Rev. Lett. {\bf 51}, 2291 (1983).
\end{thebibliography}
\end{document}